\documentclass[conference]{IEEEtran}

\usepackage{epsfig}
\usepackage{graphicx}
\usepackage{color}
\usepackage[ruled,vlined]{algorithm2e}
\usepackage{epstopdf}

\begin{document}
	
\newtheorem{defn}{Definition}[section]
\newtheorem{thrm}{Theorem}[section]
\newtheorem{prop}{Proposition}[section]
\newtheorem{lemm}{Lemma}[section]
\newtheorem{obsv}{Observation}[section]
\newtheorem{corr}{Corollary}[section]
\newtheorem{exmp}{Example}[section]

\title{\bf \normalsize \Large Smart Assessment of and Tutoring for Computational Thinking MOOC Assignments using MindReader}

\author{
\IEEEauthorblockN{Hasan M. Jamil}
\IEEEauthorblockA{Department of Computer Science\\ University of Idaho, USA\\
	jamil@uidaho.edu}
}

\maketitle

\begin{abstract}
One of the major hurdles toward automatic semantic understanding of computer programs is the lack of knowledge about what constitutes functional equivalence of code segments. We postulate that a sound knowledgebase can be used to deductively understand code segments in a hierarchical fashion by first de-constructing a code and then reconstructing it from elementary knowledge and equivalence rules of elementary code segments. The approach can also be engineered to produce computable programs from conceptual and abstract algorithms as an inverse function. In this paper, we introduce the core idea behind the MindReader online assessment system that is able to understand a wide variety of elementary algorithms students learn in their entry level programming classes such as Java, C++ and Python. The MindReader system is able to assess student assignments and guide them how to develop correct and better code in real time without human assistance.
\end{abstract}

\vspace*{2mm}


\IEEEpeerreviewmaketitle

\section{Introduction}
\label{introduction}

A significant demand is known to exist for computer science (CS) graduates, and the US government has
responded with the passing of the America Competes Act of 2007
\cite{Kuenzi2008} and subsequent refunding in 2011 to help train the much needed workforce.  Additionally, the National
Science Foundation has introduced the ``CS for All" program
with the goal that ``all students should have the opportunity to learn
CS in school."  This imperative requires that CS
education move into K-12 poorly funded schools with woefully under prepared staff
to provide education in this field. There are
few teachers with the skills necessary to teach CS courses.  In rural
schools the situation is exacerbated by the fact that there may only
be one teacher with math or science skills for the entire school.

Technological advances and economic realities are also prompting a shift in the way we learn, teach and deliver instructions to train our labor force. Tech savvy younger generation today find personalized online systems engaging and useful and are welcoming online and digital learning in all three settings -- formal or institutional, blended, and self-paced and non-formal learning spaces. The expectation is that online systems will overcome much of the hurdles we face in formal education systems and will complement it in a larger way. Although some skepticism exists \cite{DominguezESBH16,Sayapin13s}, the excitement around Massive Open Online Courses (MOOC) and more institutional approach to digital learning using BBLearn\footnote{http://www.blackboard.com/} that are at the two ends of a spectrum, are fueled by these promises. All online universities, academies and institutes such as Coursera\footnote{https://www.coursera.org/}, KhanAcademy\footnote{https://www.khanacademy.org/computing/computer-programming}, and MITOpenCourseware\footnote{https://ocw.mit.edu/courses/intro-programming/} are then immediately faced with problems in three axes -- content delivery, teaching and tutoring and assessment, much like the traditional systems do. They also grapple with enrollment and coverage, retention, cost, teaching effectiveness, and so on much like their formal counterparts. To combat these problems, new learning environments such as immersive, game-based, blended, personalized, self-regulated and self-paced, social, peer, and pair learning have been proposed, the effectiveness studies of which are ongoing \cite{EdwardsSF10,Farag12s,Sharp16s}.

But what we anecdotally know already are of significant concern. For example, the retention rate in first year programming classes is extremely low nationally. A recent online study \cite{Holton2016s} found that about 60\% STEM subject students drop out or transfer and about 55\% never graduate in state and community colleges. MOOCs and other online institutes' retention rates are even worse -- about 90\% enrollees never complete their courses \cite{Holton2016s,Tauber2013s}. We believe the environment that currently exists within the online education community does not support many of the recommendations of experienced educationists summarized in reports such as \cite{Holton2016s,Jazzar2012s}, and appear to retain the drawbacks of the traditional systems, and offer a mixed mode hodgepodge or a ``succeed on your own" online setting.

However, the encouraging fact is that there have been significant progress in several areas of computer science that we believe can be leveraged and assembled together to build effective and smart cyber systems for online teaching, tutoring and assessment of entry level computing classes, and other STEM subjects. In our vision, such a system will complement a human instructor or mentor, and take on the role of a human observer to monitor students in real time and detect where she is making a mistake in her coding exercise, and immediately offer assistance by providing diagnostic comments and helpful pointers that most likely will cure the error \cite{MartinPSR17s}.

In this paper, we introduce a novel prototype online system for tutoring and assessment, called the {\em MindReader}, for high school and freshman college students to aid learning programming languages. We develop necessary computational technologies to advance the science of computer program understanding needed for digital tutoring, and online real time assessment of programming assignments. This system is aimed at complementing human instructors at a more massive scale fully automatically. For the want of space and brevity, however, we highlight only the salient features of MindReader and refer readers to \cite{MindReader-Tech-2017} for a more detailed discussion.

\section{Related Research}

While some progress has been made in online instruction delivery, as well as in creating exciting learning environments, real time assessment and tutoring of STEM subjects online still remain at its infancy. Most often than not, these two areas rely largely on human interaction or MCQ tests, effectiveness of which are still being debated in general \cite{Simonova14s}, and in STEM settings \cite{Azevedo15s} and for computer programming classes \cite{ShuhidanHD10s} in particular. The general consensus appear to be that MCQ tests are great tools for formative and diagnostic assessments but for summative assessment, tests such as authentic assessment are more appropriate, particularly in computing courses.

The challenges in designing smart cyber systems for tutoring and summative assessment are manifold. Ideally, a tutoring or assessment system should not rely on a specific procedure for establishing correctness of a proof in mathematics, for example. Rather the logical argument in any order must be the basis. Such an assumption rules out most of the current approaches to establishing correspondence of a student response to a known solution. A few automated systems have attempted to capture this spirit in subjects such as mathematics \cite{GluzPMGV14s} and physics \cite{Mehta01s} education with extremely narrow success. In computer science, the success has been mixed \cite{NavratT14s,SorvaS15s}.

To assess programming assignments, usually understanding the code semantically is required, and for a machine it would essentially mean determining the functional equivalence of a reference solution and the student solution, which is theoretically hard -- deciding functional equivalence of two programs in general is NP-complete \cite{HuntCS80}, and only in limited instances and for special classes of programs we are able to do so \cite{EiterFTW07s,ChaudhuriV94s}. Undeterred by this weakness, researchers took a different route and tried to assess correctness of programs by various means so that the method can be used in learning exercises and online settings \cite{DrabentM05} but faced complexity barriers of a different nature \cite{Hungar91s}. Other approaches used test data to assess correctness \cite{TangSRW16s,LiYS16s} to match with known outcomes and ``assume" correctness. We, however, are not aware of a system capable of tutoring or assessing computer programs fully automatically and comprehensively.

\section{CDGs: Hierarchical Concept Structure}

The {\em program dependence graph} (PDG) \cite{Gorg16} based matching approach to determine code equivalence for the purpose of grading programming assignments is too simplistic although such approaches have been narrowly effective in detecting code clones \cite{LiKKL16s,WagnerABOR16s} and plagiarized codes \cite{LiuCHY06s, ZhangW0Z14s}. In particular, such techniques call for a complete enumeration of all possible solutions for every assignment, a largely daunting task, if not impossible. For example, consider an assignment that involves writing a code segment to swap the values of two variables. As shown in figure \ref{swap}, a student cannot be penalized if she offered  the code segment as a possible solution even though a PDG based matching approach will most likely fail to accept it as a possible solution. If a student offers a more sophisticated but unanticipated solution instead as shown below, she should be assigned higher credits, not less, though a PDG based grading will certainly be ineffective.
\begin{quote}
{\footnotesize
\begin{verbatim}
void swap(int *i, int *j){
  int t = *i;
  *i = *j;
  *j = t; }
\end{verbatim}
}
\end{quote}

\begin{figure}[h]
  \begin{center}
\scriptsize
    \begin{tabular}{|p{4cm}|p{4cm}|}
    \hline
\begin{verbatim}
#include <iostream>
using std::cout;

int main(){
  int a=27, b=43, t;

  cout << "Before " << a
    << " " << b << endl;
  t = a;
  a = b;
  b = t;
  cout << "After " << a <<
    " " << b << endl;
  return 0; }
\end{verbatim}
&
\begin{verbatim}
#include <iostream>
using std::cout;

void swap(int& i, int& j){
  int t = i;
  i = j;
  j = t; }

int main(){
  int a=27, b=43, t;

  cout << "Before " << a
    << " " << b << endl;
  swap(a, b);
  cout << "After " << a <<
    " " << b << endl;
  return 0; }
\end{verbatim}
\\ \hline
Reference Solution & Student Solution\\ \hline
\end{tabular}
  \end{center}
  \caption{Equivalent swapping code segments.}
  \label{swap}
\end{figure}

In this paper, we propose a novel and a more effective approach to matching solutions based on the idea of {\em concept dependance graphs} (CDGs) in which nodes are matched semantically as opposed to syntactic matching using PDGs. In a CDG, each node represents a hierarchically defined concept, and the graph represents the precedence relationship among the concepts. Thus, the matching of two CDGs have a much higher likelihood of determining functional and semantic equivalence of two code segments necessary for grading assignments, and offering tutoring help. We illustrate the idea using a simple problem of averaging a list of values in C++.

\begin{figure*}[ht!]
\centerline{\resizebox{.9\textwidth}{1.5in}{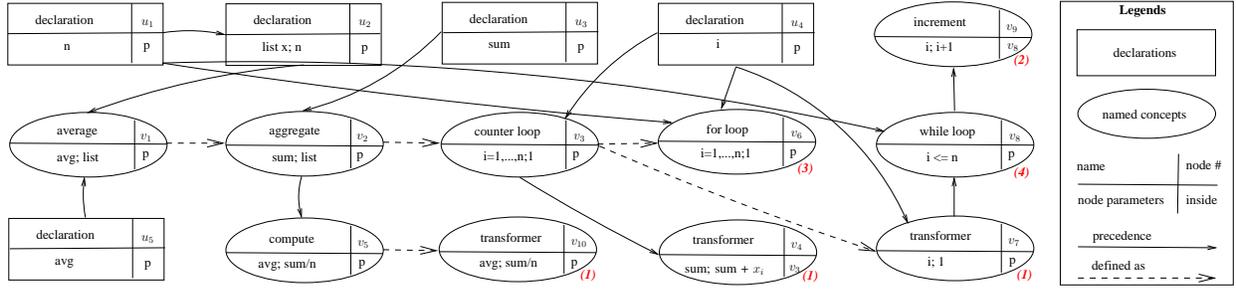}}
\caption{CDG for computing average of a list of values using counter loops.} \label{avg-cdg}
\end{figure*}

For a list of $n$ values $x_i$, their average is the simple mathematical formula $a = \frac{\sum_{i=1}^n x_i}{n}$, represented as the CDG in figure \ref{avg-cdg} in which rectangles are {\em declarations}, ellipses are {\em computable} concepts, solid {\em arrows} represent precedence, and dashed {\em arrows} represent possible replacements. In the concept symbols, there are four quadrants which represents name (upper left), contextual concept parameters (lower left), node ID (upper right), and node membership (lower right). In CDGs, concepts are defined hierarchically, and terminal nodes are either declaration, or base computable concepts such as {\em print, decide, for loop, while loop}, and so on such that all variables needed for the base concepts are also in the CDG. CDGs can be simple or complex. A simple CDG is a connected and directed acyclic graph of base concepts and declarations, while a complex CDG is a forest of simple CDGs and CDGs involving concept nodes connected using dashed arrows.

Technically, a CDG is a graph $\langle N, \prec\rangle$  of a set of nodes $N=\cup_i n_i\subseteq V$ and a precedence relation $\prec=\cup_j e_j\subseteq E$, where $V$ is the set of all possible concepts, and $E$ is all pairs $2^{V\times V}$. In figure \ref{avg-cdg}, the CDG $\langle \{u_1, u_4, v_6\}, \{u_1 \prec v_6, u_4\prec v_6\}\rangle$ is a simple, while $\langle \{u_1, u_4, v_3, v_7, v_8, v_9\}, \{u_1 \prec v_3,u_4 \prec v_3,v_3 \prec v_7,v_7 \prec v_8,v_8 \prec v_9\}\rangle$ is a complex. In figure \ref{avg-cdg}, the concept {\em counter loop} is replaceable with the sequence $v_7, v_8, v_9$, or with node $v_6$ alone. In other words, the CDG in figure \ref{avg-cdg} assumes a {\em counter loop} can be implemented in two possible ways, and thus defines an equivalence relation. Note that the concepts are also hierarchically defined. The concept {\em average} is defined as an aggregation of a list of values, followed by a division by the size of the list. An aggregate on the other hand is defined as the summation of the elements of the list inside a counter loop (note node $v_4$ is part of node $v_3$, the counter loop). Finally a counter loop is defined as a for loop or a while loop. For the student program $P_s$ below, we can transform it to construct a corresponding CDG, and match it with the conceptual solution in the knowledgebase even if the student solution is implemented using a for loop.
\begin{quote}
{\footnotesize
\begin{verbatim}
1: #include <iostream>
2: void main() {
3:   int k=0, total, size=9, mean, elements[10];
4:   while (k<=size) {
5:     total=total + elements[k];
6:     k++; }
7:   std::cout << total/(size+1); }
\end{verbatim}
}
\end{quote}

\section{Formal Model}

Let $L$ be a programming language, $\mu_L$ be a function that can parse a program $P$ in $L$ and convert each sentence into either a {\em declaration} concept or a {\em computational} concept in the language $\cal L$ of MindReader. $\cal L$ consists of two types of expressions -- {\em abstract statements} and {\em precedence relations}. Abstract statements are of two types: {\em declaration} and {\em computational} type. Declaration type expressions are tuples of the form $[N, V, T, C]$, where $N$ is the statement number in $P$, $V$ is the variable name, $T$ is the class of variable such as individual variable, boolean or a list, and $C$ is the statement or program in which the statement is included. For example, statement number 5 and 6 in the program $P_s$ above are contained in statement number 4, while the statement number 4 is contained is statement number 2. Likewise, statements 3 and 7 are contained in statement 2.

Similarly, computable expressions are tuples of the form $\langle N, E, P, C\rangle$, where $N$ is the statement number in $P$, $E$ is the type of executable statement such as assignment, loop or decision statement, $P$ is a list of context sensitive parameters, and $C$ is the statement number of which the statement is a part of. For example, statement 4 in program $P_s$ is a while loop, represented as the expression $\langle 4, whileLoop, param(cond(i<=n)), 2\rangle$, and the expression $\langle 6, tran, param(k, k+1), 4\rangle$ represents statement 6. Finally, precedence relation is a set of expressions of the form $n_1 \prec n_2$, where $n_1$ and $n_2$ are statement numbers such that $n_1$ precedes $n_2$.

\begin{figure}[ht!]
\centerline{\includegraphics[height=1.5in,width=.49\textwidth]{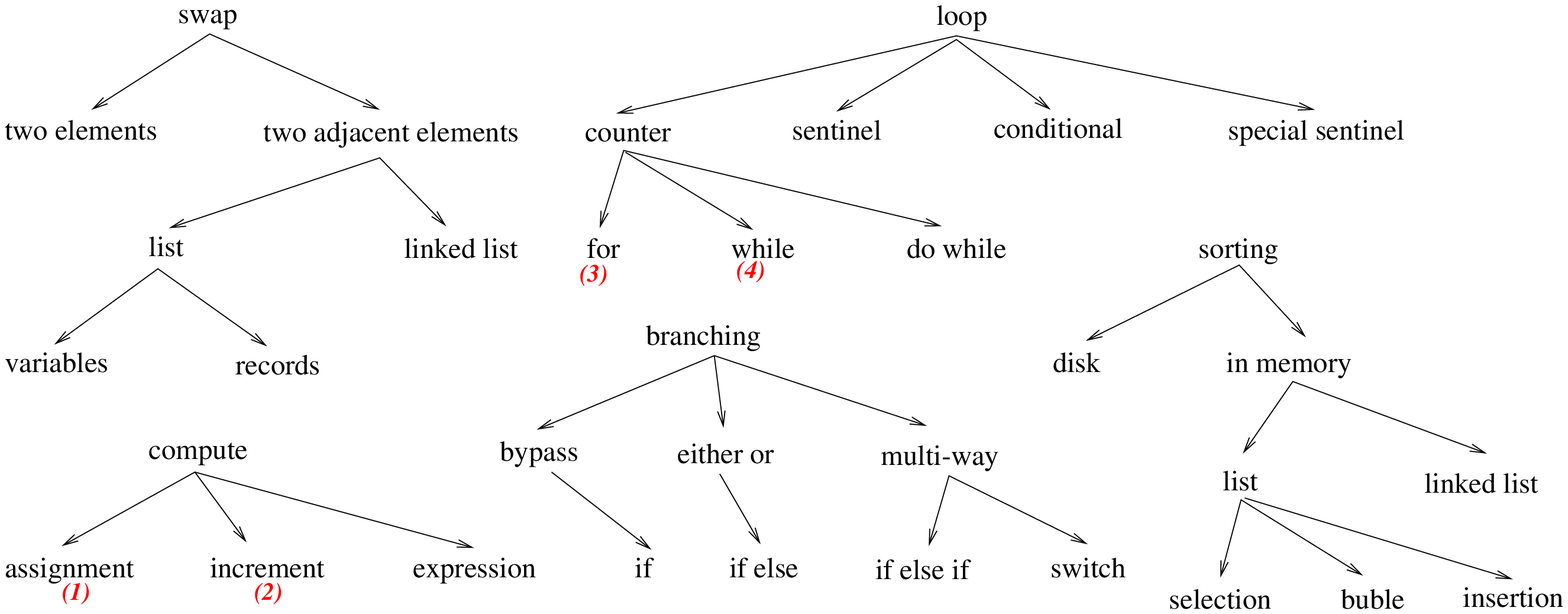}}
\caption{Example concept hierarchy.} \label{cgh}
\end{figure}

The language $\cal L$ of MindReader is a tuple $\langle \mu_L, {\cal C}, \Sigma, \Gamma, \Psi\rangle$ of a concept extractor $\mu_L$, concept hierarchy $\cal C$ (e.g., figure \ref{cgh}),  concept mapper $\Sigma$, concept dependence graph $\Gamma$, and a subgraph isomorph function $\Psi$. The concept hierarchy organizes higher level concepts from computable expressions. For example, a counter loop can be a composite of an {\em assignment}, a {\em while} and an {\em increment} statement as discussed earlier in the context of figure \ref{avg-cdg}. The $\Sigma$ function transforms the CDG created by $\mu_L$ into higher level concepts using the concept hierarchy $\cal C$ and the CDG $\Gamma$, iteratively. Therefore, given a program $P$, the least fixpoint $lfp(\Sigma(\mu_L(P),{\cal C},\Gamma))$ is the final CDG of a program $P$. Observe that the concept hierarchy $\cal C$, the CDG $\Gamma$, and the summarization function $\Sigma$ help abstract programs into CDGs and increases the matching likelihood  with high level abstract algorithms stored as a reference CDG independent of their lower level implementations.

Finally, the subgraph matching function $\Psi$ ensures proper matching of CDGs independent of the variable declarations, typing and naming. It ensures proper substitutions throughout the code segment. Therefore, for any given program $P$, if $\Psi(C_a, lfp(\Sigma(\mu_L(P),{\cal C},\Gamma))\approx 1$, where $C_a$ is the conceptual CDG of any algorithm $a$, then we assume that the submitted code is acceptable and correct. Conversely, if for any ``unknown" program $P$, $\Psi(C_a, lfp(\Sigma(\mu_L(P),{\cal C},\Gamma)))\approx 1$, then we can be confident that the unknown program is a candidate implementation of the abstract and conceptual algorithm $a$. This a significant and powerful method to determine functional equivalence of unknown codes which is extremely difficult, if not impossible, using PDG based approach due to its inability to summarize codes functionally.

\section{Assessment and Tutoring using MindReader}

\begin{figure*}[ht!]
\centerline{\includegraphics[height=1.7in,width=.9\textwidth]{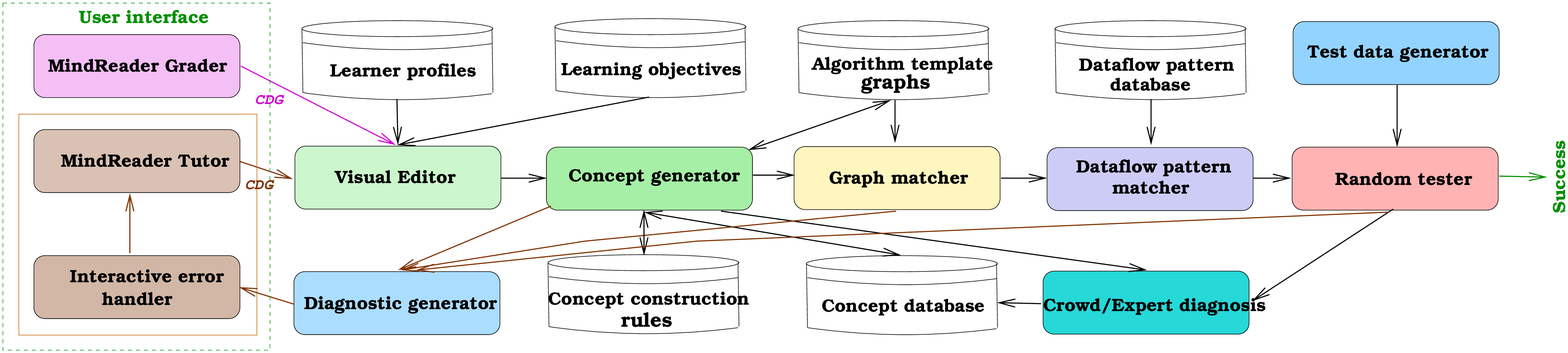}}
\caption{MindReader architecture: Brown lines for tutoring, purple line for assessment, and black lines for both system.} \label{archi}
\end{figure*}

The high level architecture in figure \ref{archi} depicts MindReader's two broad subsystems for two distinct but complementary functions -- tutoring and grading. In MindReader, all learners have a profile which includes background, past lessons, tests and tutoring activities, known problem areas, and their peer groups. MindReader generates tutorials based on students' profile and level of programming competence expected along the lines of the systems such as \cite{BoumizaSB16s} keeping in mind that for computation thinking classes, the challenges primarily involve the difficulties in learning the syntax and understanding the semantics and use of constructs such as loops, conditional statements, and simple algorithms \cite{HullsNKPB05s}. For the purpose of both grading and tutoring, MindReader assembles the statement structures written by the student into possible CDGs using the concept structures in the {\em Concept Database} according to the rules in {\em Concept Construction} rule base with the aim of matching the CDG with one of the known templates in the {\em Algorithm Templates}. Failure to match CDGs of the student code and the reference template results into a dataflow pattern match using known and random test data of the compiled codes. Failure to match flow patterns forces a diagnostic feedback, but a success indicates a new way of solving a problem unknown to MindReader, and the new CDG is included in the knowledgebase after proper curation.

\section{Learning Complex Concepts}

\begin{figure}[h]
\centerline{\resizebox{.49\textwidth}{1.25in}{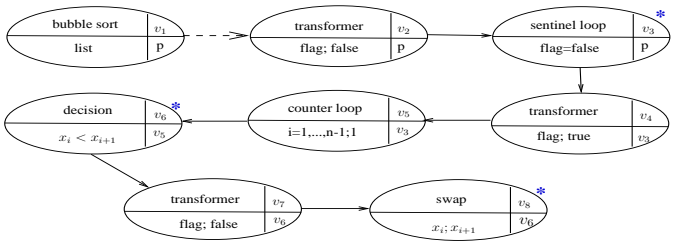}}
\caption{Bubble sort reference CDG.} \label{bubble-cdg}
\end{figure}

Building concepts hierarchically and generating corresponding CDGs though intuitive, learning new concepts could be challenging. In MindReader, we assume that it is impossible to enumerate all reference solutions regardless of the complexity. We thus adopt an incremental learning approach with the assistance of a panel of curators or experts in MindReader's architecture in figure \ref{archi}. To understand how MindReader learns new concepts, consider an abstract algorithm for bubble sort as shown in algorithm \ref{alg:bubble}, its C++ implementation $Q$ as shown below, and its CDG representation $C_b$ shown in figure \ref{bubble-cdg} as a reference solution. Obviously, the $lfp(\Sigma(\mu_L(Q),{\cal C},\Gamma))$ will not match with $C_b$, i.e., $\Psi(C_b, lfp(\Sigma(\mu_L(Q),{\cal C},\Gamma))) << 1$, since the loop in statement 1 is not a sentinel loop. But a dataflow analysis and random data test comparison will show a match, prompting a curation step and learning the rule that bubble sort can also be performed with an outer counter loop, and a reverse inner counter loop. Note that the blue starred nodes in the CDG in figure \ref{bubble-cdg} will also need to be implemented.

\IncMargin{1em}
\begin{algorithm}[h]
{\footnotesize
\KwIn{A list of $n$ values in random order}
\KwOut{Ascending order list}
set $sorted=false$\;
\While{not sorted}{
set $sorted=true$\;
\ForEach{element $i=1,\ldots, n-1$}{
\If{element $i$ $<$ element $i+1$}{
swap elements $i$ and $i+1$\;
set $sorted=false$\;
}}}
\caption{Bubble sort}
\label{alg:bubble}
}
\end{algorithm}
\DecMargin{1em}

{\footnotesize \begin{verbatim}
0: void bubbleSort(int ar[]) {
1:   for (int i = (n - 1);
                        i >= 0; i--) {
2:      for (int j = 1; j = i; j++) {
3:         if (ar[j-1] > ar[j]) {
4:              int temp = ar[j-1];
5:              ar[j-1] = ar[j];
6:              ar[j] = temp;
   }  }  }  }
\end{verbatim}}

\section{Summary and Future Research}

In this paper, we reported a late breaking result of a research focusing on semantic understanding of student codes in an online learning environment. We have demonstrated that CDG based matching code fragments have a higher likelihood of detecting semantic and functional equivalence of two programs. The process is complemented by a dataflow analysis and random testing regime to identify possible valid solutions and learn new rules. We have also demonstrated that detecting code clones and plagiarized codes based on PDGs is fundamentally different from matching two codes functionally using CDGs. In CDGs we substitute equivalent nodes under the guidance of a template CDG, and concept hierarchy to determine semantic similarity essential for grading tasks of MOOC student assignments. It should be evident that summarization of concepts in the concept graphs allows for abstract algorithm development, and it should be possible to actually write codes in various languages as an inverse function and develop new languages such as Scratch.

Initial evaluations of MindReader was encouraging and a more serious performance analysis and comparison with existing systems is being planned. Once deployed, and students use it for a period of time, we plan to collect a large number of coding examples and investigate students learning behavior, and effectiveness and do comparative analysis with the traditional classroom teaching. Identification of problem areas of learning where a significant number of students are having difficulty manifested by their inability to solve problems could imply gaps in instruction delivery, course content design or learning habits warranting a revision, and could help develop personalized teaching, tutoring and assessment regimes, and measured for continuous improvement.

\bibliographystyle{IEEEtran}

\end{document}